# LASER-INDUCED BREAKDOWN SPECTROSCOPY FOR REAL TIME AND ONLINE ELEMENTAL ANALYSIS

Virendra N. Rai[1], Awadhesh K. Rai[2], Fang-Yu Yueh and J. P. Singh

Diagnostic Instrumentation and Analysis Laboratory

Mississippi State University,

205 Research Boulevard, Starkville, MS 39759-7704, USA

1 Centre for Advanced Technology, Indore-452 013 (India)

2 G. B. Pant University of Agriculture and Technology, Pantnagar-263 145 (India)

## ABSTRACT

Laser-induced breakdown spectroscopy (LIBS) is a laser based diagnostics used to study atomic emission from the expanding plasma plume formed during the laser-matter interaction. It provides valuable information about the composition of the target material. LIBS has proved its potential application in the analysis of impurities, pollutants and toxic elements in various types of matrices of different samples (solid, liquid and gases), even those present under difficult and harsh environmental conditions. This article reviews some recent developments in the field, and its wide application in various fields of research and analysis.

## 1. INTRODUCTION

The interaction of high-power laser light with a target sample has been an active topic of research not only in plasma physics but also in many field of research and analysis [1]. The use of lasers to vaporize, dissociate, excite or ionize species on material surfaces has the potential of becoming a powerful analytical tool. When a high-power laser pulse is focused onto the target of any kind of material (solid, liquid and gases) the irradiation in the focal spot can lead to rapid local heating, intense evaporation and degradation of the material. The interaction between a laser beam and the material is a complicated process dependent on many characteristics of both the laser and the target material. Numerous factors affect ablation, including the laser pulse properties, such as pulse width, spatial and temporal fluctuation of the pulse as well as the laser power fluctuations. The mechanical, physical and chemical properties of the sample target also influence the ablation process. The hot laser-produced plasma radiates various types of emissions, ranging from x-rays to visible emissions. The spectroscopic study of line emission from the micro plasma (Atomic Emission Spectroscopy) can provide an information about the composition of the material. This technique based on the spectroscopic study of optical emission from laser produced plasma is popularly known as laser-induced breakdown spectroscopy (LIBS). LIBS is an atomic emission technique suitable for quick and on-line elemental analysis of any phase of material and has proved its importance in obtaining analytical atomic emission spectra directly from solid, liquid, and gaseous samples [2-5]. It has been used successfully in the analysis of trace elements present in solid, liquid and gaseous samples as well as in the detection of radioactive elements [4]. The sensitivity of this system is an important factor, which requires an improvement for the detection of a minor or trace elements in the samples. Various techniques have been utilized to enhance the sensitivity of LIBS system; that is, emission intensity coming out from the laser produced plasma. This includes oblique incidence of laser on the sample surface, double pulse excitation and the introduction of purge gas around the microplasma etc. Magnetic field has also been used to enhance the emission from laser induced plasma in different experimental conditions [6]. Plasma changes its various physical properties during expansion across the magnetic field, which ultimately affects its emission characteristics. Enhancement in emission from the plasma in the presence of magnetic field occurs mainly as a result of confinement of the plasma, where the kinetic energy of plasma is transformed into the thermal energy, which helps in heating the plasma [7,8]. Double pulse excitation technique has also been found applicable in enhancing the emission from laser produced plasma. During this experiment, the first pulse generated a pre-formed plasma, which is further excited by the second laser pulse after few microsecond time duration, which reheats the pre-formed plasma [9,10]. It has been shown that the volume of plasma formed after second laser pulse is about twice as large as that formed with a single laser. Along with increase in volume of emitting plasma an increase in plasma temperature as well as increased ablation also contributed towards enhancement in plasma emission. LIBS has various advantages over the conventional laboratory based techniques. It is a sensitive optical technique with high spatial resolution (small focal spot). In this process vaporization and excitation of the sample materials occurs directly in one step. In fact either no or little sample preparation is needed and only a small amount of sample is required, which saves the time of sample preparation. Finally the beauty of this technique is that it can provide an on-line and real time analysis of the samples from remote distance, which requires no direct contact with the sample.

The primary aim of this article is to demonstrate the applicability of the LIBS process in the elemental analysis of solid, liquid and gaseous samples along with an enhancement in its sensitivity.

## 2. LIBS OF SOLID SAMPLE

LIBS can be used to study and analyze both conducting and non-conducting solid samples. As the laser beam is focused on solid surface, the sample material absorbs the laser energy to melt and vaporize a certain amount of material. The vapor absorbs laser energy and forms a high temperature plasma near the sample surface. The plasma expands into the atmosphere and transfers its energy in various forms. The emission lines from the highly ionized ions can be found close to the target surface, whereas emission from the singly ionized and neutral particles appear further away from the surface. Since the laser breakdown threshold is lower in solids than in liquid and gaseous samples. Generally less energy is needed for solid sample application. More papers describing analytical results of LIBS on solids were found than the papers dealing with either liquids or gases. Several publications are available in literature describing the elemental determination in steel, Al alloys, soil, and paints, using the LIBS technique [4]. LIBS has also been used for on-line quality control of rubber mixing and for analyzing mining ores. The review by Rusak et al. [3] and Yueh et al. [4] reports various applications of LIBS in the study of solid samples. It is found that LIBS is a most suitable technique for field-based industrial applications, which can provide a real-time and on-line analysis of material for process control and monitoring. Most of the experimental methods reported earlier are laboratory systems, where plasma is generated by focusing the high intensity laser beam on the sample surface using an assembly of lenses. The light emitted from the plasma is collected either by the same assembly of lenses or different assembly and is focused on the entrance slit of the spectrometer for further analysis. Such an experimental setup may not be well suited for field measurements. A field system usually requires minimal and flexible optical access to the test facility and minimal on-site alignment.

A simple and robust fiber-optic (FO) probe has been developed presented that uses one optical fiber both for delivering the laser power to produce a spark as well as for collecting the resulting emission from the spark for quantitative elemental analysis with greater accuracy and a lower detection limit [11]. A schematic diagram of the FO LIBS probe is shown in Figure 1. The second harmonic (532 nm) of a pulsed Nd: YAG laser (Big Sky, Model CFR 400) operating at 10 Hz (pulse duration 8 ns, beam diameter 7 mm, and the full angle divergence 1.0 m rad) is directed toward the optical fiber using a harmonic separator and a 532-nm dichroic mirror. A specially coated ($45^{\circ}$ angle of incidence) dichroic mirror (DM), which reflects at 532 nm and transmits the spectral range from 180-510 nm and 550-1000 nm, was used to reflect the laser beam and to pass the LIBS signal to the optical fiber. The laser beam transmitted through the optical fiber is collimated using a 10-cm focal length lens and then focused on the sample with the help of a 5-cm focal-length lens. The same set of lenses and optical fiber also collect the emission from the laser-produced plasma around the focal point. The collimated emission then passes through the dichroic mirror and is focused onto a different optical fiber bundle (round-to-slit type) with a 20-cm focal length lens. The slit-type end of the fiber bundle delivers the emission to the entrance slit of a 0.5-m focal length spectrometer (Model HR 460, JOBIN YVON-SPEX) equipped with a 2400-lines/mm grating blazed at 300 nm. A gated intensified charge couple device (Model ITE/CCD, Princeton Instruments) was used as the detector with its controller (Model ST 133, Princeton Instruments). A programmable pulse delay generator (MODEL PG-200, Princeton Instruments) was used to gate the ICCD. The entire experimental apparatus was under the control of a computer (Dell Dimension M 200a) running the WinSpec 3.2 software (Princeton Instruments). Multiple (100) spectra were averaged to get one spectrum. Fifty such spectra were recorded and stored in one file for analysis to get an average area/intensity value for the intended line.

LIBS spectra of different Al alloys were recorded. The intensity of the lines was measured by integrating the peak area with an automatic spectral baseline correction. The quantitative spectral analysis involves relating the spectral line intensity of an element in the plasma to the concentration of that element in the target that provides a calibration curve. We analyzed the most important minor elements in Al alloy, which include copper, magnesium, manganese, nickel, chromium, and iron. It is well known that strong continuum emission is generated at the beginning of plasma formation. The plasma then cools by expansion processes. The measurement in the first few microseconds is hence unsuitable for obtaining the calibration curve. Normally LIBS data were recorded at 2-μs gate delay, which was used for calibration purpose. A typical LIBS spectrum recorded using above experimental parameters is shown in Figure 2. After optimizing the experiment, we obtained calibration curves by using different aluminum alloy samples having different concentrations of minor elements. The calibration data were collected under the same experimental conditions for each sample. The analyte lines of Cu, Cr, Mn, Fe, and Zn, were used to obtain the calibration curves, shown in Figures 3 and 4. Each figure relates the emission line intensity of an element to its concentration in the aluminum alloy.

## 3. LIBS OF LIQUID SAMPLES

The detection and quantification of light and heavy elements within liquid samples are important from application points of view, particularly in industrial processing, environmental monitoring, and waste treatment. Golovlyov and Letokhov [12], Esenaliev et al. [13], and Oraevsky et al. [14] have studied the physical mechanisms of the laser-induced breakdown on liquid samples. Initially, liquid samples were studied by focusing the laser in the bulk liquid, which creates heavy splashing as a result of shock waves [15-17]. These effects change the position of the liquid surface with respect to the laser focus, which adversely affects the analytical results. Instead, generation of plasma in the liquid prevents splashing if the plasma is formed on the liquid surface. The bulk generation of the plasma presents another drawback in terms of a decrease in the duration of plasma emission; that is, plasma light observation time is extremely short, usually of the order of 1 μs or less. Cremers et al. [17] found in their experiments on liquid samples that, in general, plasma parameters (plasma density and its temperature) could not be derived accurately for delay times beyond 1.5 μs.

The work described in this section stems from a need in the nuclear industry to conduct real-time on-line analysis of radioactive waste (technetium Tc) in liquid specimens. These are encountered during the reprocessing of nuclear fuel and in monitoring of nuclear waste present in the storage tanks. Technetium is a radioactive element and a product of the nuclear power cycle. The most stable Tc isotope has a half-life of $2.1\times10^5$ years and decays via beta emission. Due to the long half-life and the relatively high yield from uranium decay, it is desirable to separate technetium from nonradioactive and short-life components found in the tank waste. It is also important to isolate it with other long-life radionuclides in geologically stable waste for long-term safe storage. Similar problems are also encountered in other industries where toxic liquid effluent and/or waste are present, and a real need exists for proper regulation of these materials. This requires a real-time, remote, on-line LIBS analysis system. Enhancement in the sensitivity of LIBS system for the study of liquid samples is also required. Sensitivity of LIBS system is mainly dependent on the emission intensity from laser produced plasma, which needs to be increased. Two techniques were tried to increase the emission from the plasma of liquid samples. These are the generation of laser produced plasma in the presence of steady magnetic field (0.5 T). In this case enhancement in intensity occurred due to confinement of plasma. In another technique double laser pulse was used to excite the plasma.

The schematic diagram of the experimental setup for recording the laser-induced breakdown emission on the bulk liquid surface as well as in the case of a laminar jet is shown in Figure 5. A Q-switched frequency doubled Nd: YAG laser (Continuum Surelite III) that delivers energy of ~400 mJ in a 5-ns duration was used in this experiment. The laser was operated at 10 Hz during this experiment and was focused on the target (in the center of the liquid jet or on the surface of the bulk liquid, depending on the experiment) using an ultraviolet (UV) grade quartz lens with a focal length of 20 cm. The same focusing lens was used to collect the optical emission from the laser-induced plasma. The data collection/ acquisition system is similar to that used for solid sample described in previous section.

The general configuration used for LIBS on solid samples (laser beam perpendicular to the solid surface) leads to splashing in the case of liquids. Splashing results in covering the focusing optics with droplets and therefore prevents further quantitative use of this technique. This was mainly due to the fact that the plasma expansion is directed perpendicularly to the target surface. The LIBS of liquid jet has various other advantages in comparison to bulk liquids [5]. A Teflon nozzle of diameter ≤1-mm was used with a peristaltic pump (Cole-Parmer Instrument Co.) to form a laminar liquid jet. The laser was focused on the liquid jet such that the direction of laser propagation was perpendicular to the direction of the liquid jet. The laser was focused ~15 mm below the jet exit, where the liquid flow was laminar. However, the extent of laminar flow depended on the speed of the pump. The liquid jet was aligned in a vertically downward direction.

For quantitative measurement, the measured emission intensities should be related to the absolute or relative elemental concentration. The system response must be calibrated for a certain measurement. To calibrate LIBS system for liquid measurement, system was optimized for different atomic and ionic lineemission by adjusting the gate delay time and gate width of the detector, as well as the laser energy. The LIBS signals of the various seeded elements (Cr, Mg, Mn, and Re) were recorded at different sample concentrations to obtain the calibration curves at optimized experimental conditions for estimating the limit of detection for the elements in the liquid sample. Figure 6 shows the calibration curves for rhenium (Re) obtained from liquid-jet measurements at a delay time of 8 μs and a gate width of 15 μs at two different laser energies. This shows that the LIBS signal increases linearly with concentration. An increase in excitation laser energy increases the LIBS sensitivity for each concentration. However, the factor of increase in the LIBS signal varies from sample to sample. The detection limits for Cr, Mg, Mn, and Re were calculated based on the calibration data.

The detection limit for the seed elements of Cr, Mg, Mn, and Re were obtained as 0.4, 0.1, 0.87, and 10 μg/ml, respectively [6]. The LOD of various elements such as Pb, Si, Ca, Na, Zn, Sn, Al, Cu, Ni, Fe, Mg, and Cr obtained from bulk water and oil matrices has been reported in the literature [18]. Similarly, the LOD of elements Al, Ca, Cr, Cu, K, Li, Mg, Mn, Na, Pb, Tc, U and Re were obtained using the liquid-jet system (Table 1)[5]. The LOD of Cr, Mn, Mg, and Re obtained in this experiment was found better in comparison to the results described in the literature. It was noted that the emission intensity of plasma from Mg, Mn, Cr increased by nearly ~1.5 times in the presence of steady magnetic field of 0.5 T. The measurement of limit of detection of magnesium in the presence of steady magnetic field showed a decrease by a factor of nearly two, that is, from 1.74 in the absence of magnetic field to 0.83 ppm in the presence of magnetic field [8]. An increase in plasma emission by nearly six times was obtained from double laser pulse excited Mg and Cr plasma. The limit of detection measurement of chromium showed nearly an order of magnitude decrease, that is from 1300 ppb in single pulse excitation to 120 ppb in double pulse excitation [10]. This indicates that both the technique such as use of steady magnetic field and double pulse excitation enhances the sensitivity of LIBS. However double laser pulse excitation was found more effective. Various experimental parameters such as matrices, wavelength of emission, gate delay, and the process of obtaining a calibration curve affect the estimation of the limit of detection. Which is why an exact comparison of the LOD data from the two research teams is difficult.

The limit of detection reported for various elements in this experiment as well as in the literature has proved the LIBS technique suitable for finding pollutant (trace) or minor elements at high and moderate concentrations. However, it is not possible to detect them at very low concentrations. Still a serious effort is needed to make the system more versatile for very low concentration measurements.

## 4. LIBS OF GAS SAMPLES

Laser-induced breakdown in gas has been studied extensively [19]. The breakdown thresholds in the atmospheric pressure gases are proportional to the ionization potential of each gas divided by the collision frequency. Typically, it needs a laser power density that corresponds to an electric field strength of the order of $10^5$ volt/cm to produce breakdown in the gas pressure around 1 atm. Due to the presence of micron-sized aerosols and impurity particles, the observed laser breakdown threshold in gas samples is generally lower than that from the theoretical prediction. This is because the particles acting as seeds can significantly lower the breakdown threshold of clean gas. Typically, laser-induced air breakdown has a plasma temperature of 20,000 K and an electron density of $10^{17}$-$10^{18}$ $cm^{-3}$ after the plasma is formed.

The application of LIBS for gas analysis involved a focused high-energy pulsed laser to produce the breakdown. The emission from the laser-induced plasma can be used directly to measure the composition of gas, eliminating the need for sample preparation. Schmieder et al. were first to show that LIBS could be applied as a combustion diagnostics for monitoring the elemental constituents of a combustion product [20, 21]. They used a time-integrated photographic technique and diode array to detect N and O in gas mixtures and to measure the C/N ratio in the flame. Radziemski and Loree pioneered LIBS applications on gas measurements using time-resolved detection [22]. They used a time-gated optical multichannel analyzer or a PMT-boxcar detection system and found the detection limits for P and Cl in air as 15 and 60 ppm, respectively [22]. Cremers and Radziemski were later able to detect Cl and F in air with a detection limit of 80 ng and 2,000 ng, respectively [23]. They also found that the absolute detection limit for Cl and F could be improved in a He atmosphere. Radziemski et al. have used LIBS to detect Be, Na, P, As, and Hg also in air [24]. Due to the small sample volume and possible sample inhomogeneity, LIBS' measurement precision in a gas sample is generally poor. The various size particulates in the gas can cause the breakdown to be generated at different locations along the axis of the laser beam and lead to significant signal variations. The most common interference found in the air breakdown is due to the CN emission. CN is produced from the reaction of C and N, which are produced in the spark. The intensity of CN bands depends on the concentration of a C-containing compound in the gas stream. The analyte lines in the CN band covered spectral region have less sensitivity due to the spectral interference.

The characteristics of LIBS in gas measurements have been discussed in detail and can be found elsewhere [25, 26]. In this section we only discuss calibration techniques and various applications for gas samples developed at DIAL using a versatile, mobile LIBS system. This system was originally developed to monitor toxic metal concentrations in the off-gas emission of a plasma hearth process system. It has been used to conduct various laboratory studies and field measurements for different applications.

The experimental arrangement of the LIBS system for gas samples is similar to that used in the liquid experiment, described in the previous section. The laser beam was directed and focused on the desired gas sample with a 10-cm or 20-cm focal lens. The emission from the spark was collected using an UV optical fiber bundle and sent to the detection system. In some cases, we used two detection systems to monitor two spectral regions simultaneously or two different measurement locations. The second detection system includes a SPEX 500M spectrograph equipped with a 1024-element intensified diode array detector (Model IDAD-1024, Princeton Applied Research). A fiber bundle with the output end of the optical fiber bundle splitting into two bundles was coupled to two spectrographs for the measurements of two spectral regions. A notebook computer (Model T-4700CS, Toshiba) was interfaced to the second detector controllers for data acquisition and analysis. To maximize the signal, a gate pulse delay of 5-10 μsec and width of 10 to 20 μsec was used in most of the work Since only one optical port was required in this configuration, it was suitable for real-time measurement in off-gases.

To quantify the LIBS data of gas sample, we need to calibrate LIBS instruments with the known concentration samples. The gas phase sampling generally uses a nebulizer to produce aerosol from the solution of standard reference materials. We used an ultrasonic nebulizer (USN) (Cetac U-5000AT$^+$) to produce the dry aerosols of selected metals. Two setups were used for LIBS calibration, as shown in Figure 7. Calibration was performed by injecting known concentrations of dry aerosols from an ultrasonic nebulizer into either a sample cell (closed system) or air (open system). Volumetrically diluted plasma emission standard solutions (Spex Industries) were injected into the USN with a peristaltic pump at a rate of 1.9 mP/min. A 0.8-P/min flow rate of air was used as a carrier gas flow to transport the aerosol through the USN. The aerosol in the USN was first dried by a heated (140 EC) tube and then passed through a chilled (3 EC) condenser to remove water. In the open system, the dry aerosol from the USN was sent to a stainless steel sample injection tube, and the laser beam was aligned 2 mm above the end of the tube and focused on the center of the tube to achieve reliable calibration. The sample injection tube was enclosed in a Pyrex cylinder to reduce interference from the surrounding air. In the closed system, the metal aerosol was injected continuously to the sample cell. The sample cell is made of PVC. LIBS calibration data were collected after the composition equilibrium in the cell was reached. The waste solution was collected during the nebulization procedure. The USN system was later operated with the collected waste solution. In the laboratory, we used both the open and closed system to calibrate the LIBS system. In the field measurement we used the open system to conduct on-site calibration. The on-site calibration was generally performed before the field test started and after the test ended each day to verify the system response. The on-site calibrations were performed by injecting metal aerosol generated from an USN into the gas stream using a probe. The sample injection probe was mounted on the opposing port across the gas stream. Each day, the LIBS spectra were also recorded before the metal injection for zero check.

Calibration is the most difficult issue in the development of LIBS, especially for the field measurement. LIBS is an atomic emission spectroscopy. The parameters, which can affect the characteristics of the LIBS spectra in gas, include particle size, gas pressure, temperature, and laser energy. For quantitative measurement, the calibration procedure should keep those parameters as close to those in the measurement as possible. For on-site measurement, LIBS is considered a non-sampling technique. However, this implies an extra difficulty for calibration.

An extensive LIBS calibration study has been performed [27, 28]. Two calibration methods, a hydride generator and an USN, were compared. LIBS spectra were recorded using a hydride generator (see Figure 8) and an USN with a mixture solution of As, Sb, and Sn in $N_2$ and He to study interference effects among different metals. Since HCl concentration plays an important role on hydride generation efficiency, different HCl concentrations in the mixture solution were also used in this study. The spectral interferences were not significant in this study. However, the results from the hydride generation were quite sensitive to the acid concentration in the mixture. Comparison of metal generation from metal oxide particles produced by an ultrasonic nebulizer shows that the actual gas stream metal distribution is close to that from the USN. Efficient metal hydride generation requires different acid concentrations for different metals. An USN, on the other hand, is ease to use, and works for all resources conservation and recovery act (RCRA) metals. Therefore, the calibration curves for every RCRA elements were obtained using an USN. Based on the data collected from the USN and, after averaging 50 laser pulses, the precision for most of the RCRA metals was 10%, and the accuracy was 5-10%. Studies of relative calibration were also performed to implement on-line calibration in field measurements.

Based on the experimental results, we decided to use the USN to conduct the LIBS calibration for gas samples. We have used the calibration data obtained by injecting known concentrations of dry aerosols from the USN into air. Generally, LIBS data from four concentrations of an element were used to obtain the calibration curve. The calibration curve is based on either peak height or peak area of each analyte line. The slope of the calibration curve is used as the calibration factor to infer metal concentration. The peak area (or peak height) of an analyte line from a demonstration test on LIBS spectra was normalized using its calibration factor to obtain the metal concentration. In general, peak height calibration and peak area calibration give about the same result for an interference-free line. For different types of spectral interferences, either peak height or peak area must be selected for best results. We found that the peak area analysis yielded better results than did the peak height analysis for the self-absorbed spectral line, and the peak height analysis yielded better results than did the peak area analysis for a line overlapped with other lines.

The limits of detection (LOD) of selected analyte lines of seven RCRA metals determined at the DIAL laboratory just before the CEM test are listed in Table 2. The precision and accuracy of these measurements are estimated from the calibration data and are also listed in Table 2. The precision and accuracy greatly depends on laser pulse-to-pulse fluence fluctuations at focal volume and the concentration variation in the aerosol flow from the ultrasonic nebulizer. The accuracy and precision of LIBS measurements can be improved by increasing the signal integration time. Since some lines have spectral interferences, the actual field detection limits may be slightly higher than the reported detection limit, depending on the concentration of the interfering elements. The LODs depend on the experimental conditions and can be reduced by improving the optical design and detection system.

In some cases, if the absolute concentration calibration is too difficult to obtain due to the variation of the environmental conditions, relative concentrations may be considered. One can either use the calibration based on the intensity ratio of the analyte line and reference line or fit the observed spectra with the theoretical model [29]. Analysis of LIBS data using spectral fitting requires the knowledge of spectroscopic constants such as plasma temperature, and degree of ionization. These two parameters, however, are not easy to be determined accurately. Alternately, D. K. Ottesen et al. used reference line intensity information from the NIST collections to perform spectral fitting to obtain relative concentration [30]. However, the excitation condition needs to be checked for this simple method since the intensities in these reference collections are obtained under certain conditions, which may be very different from induced plasma.

Ciucci et al. have recently developed an algorithm for calibration-free quantitative LIBS analysis and seem to have had great success with laboratory data [31]. However, such an approach relies on some basic assumptions, such as laser-induced plasma (LIP) in LTE existence; LIP is an optically thin plasma composition representative of the actual sample composition, etc. It should be evaluated with the practical data, because environmental conditions fluctuate over time.

The practical environments are quite different from those in a laboratory. Transferring the LIBS calibration obtained in a laboratory to field measurements is a great challenge. To establish a calibration scheme for quantitative measurement in practical environments, we conducted a series of studies to correlate LIBS backgrounds with changes in excitation conditions. A linear relationship between the LIBS calibration slope and the backgrounds for Cd and Be was found. These data were obtained from spectra recorded in the 230-nm spectral regions with different laser energies, gate windows, and test cells. Figure 9 shows the linear relation between the laser energy and the Cd calibration slope. The LIBS background was also found to be linearly proportional to the calibration slope (see Figure 9). The data were recorded with gate delays of 20 μs and 15 μs with a fixed gate width of 30 μs. These results imply that the background can be used to correct the changes in plasma conditions. However, the same experiment in the 415-nm spectral region shows a linear relationship between background and calibration slope present only when laser energy is below a certain limit (see Figure 10). At higher laser energy, the CN interference is dominant in this spectral region, and the intensities of the analyte lines of Pb and Cr are possibly saturated. The results of the background study show that background normalization can be used to correct the calibration factor due to minor changes in the plasma condition. However, this approach demands great care due to its limitations.

**CONCLUSION**

Considerable progress in the area of basic and applied research of LIBS has been noted during the last two decades due to an improvement in several experimental parameters to detect trace elements in solid, liquid, and gaseous samples. Many research laboratories all over the world are working in this field.

In this article, we have presented the applicability of LIBS in the analysis of solid, liquid, and gaseous samples. It demonstrates that LIBS may be useful for elemental analysis of different types of solid samples (steel, Al alloy, sand, and biological tissue). For quantitative measurements of solid samples in different industries, a calibration method must be developed to overcome matrix effects, pulse-to-pulse plasma fluctuation, sample-to-lens distance effects, and self-absorption problems. Currently, no real-time measurement of melt composition is available for the aluminum, steel, and glass industries. To improve the production efficiency, a technique that can provide rapid, in-situ melt composition measurements is needed for these industries. This way LIBS has great potential application for molten material. Work is in progress for the application of LIBS in analyzing the liquid samples to reduce the LOD of trace elements particularly for making a technetium monitor. An application of steady state magnetic field, double laser pulse excitation and the use of purge gas around the liquid jet have already showed an improvement in the sensitivity (max. by an order of magnitude) of the LIBS system. However further improvement is required. The combination of a steady magnetic field with another signal-enhancing technique, such as the use of double-pulse excitation and purge gas experiments, must be tried in order to further improve the LOD of trace elements in liquids, an endeavor that will be advantageous particularly for nuclear industry. LIBS sensitivity needs to be improved even for continuous emission monitoring of trace and toxic elements present in the off gases of any combustion systems. More work is also required to improve the calibration methods. Specifically, an on-line calibration method needs to be developed for CEM.

## ACKNOWLEDGEMENTS

This work is supported by U.S. Department of Energy Cooperative Agreement No. DE-FC26-98FT 40395.

**RFERENCES**

Table 1. Limit of detection of elements recorded in liquid-jet experiment.

| Elements | Wavelength (nm) | Detection limit (ppm) [Ref # 6] | Limit of detection (ppm) |
|---|---|---|---|
| Al | 396.15 | 18 | |
| Ca | 422.67 | 0.6 | |
| Cr | 520.45 | 200 | |
| | 425.43 | | 0.4 |
| Cu | 324.75 | 5 | |
| K | 766.49 | 4 | |
| Li | 670.77 | 0.009 | |
| Mg | 285.21 | 3 | 0.1 |
| Mn | 403.08 | 10 | 0.87 |
| Na | 588.99 | 0.08 | |
| Pb | 405.78 | 40 | |
| Tc | 429.71 | 25 | |
| U | 409.02 | 450 | |
| Re | 346.05 | | 10 |

Table 2. Limit of detection of some selected metals in gas.

| Element | Analyte Line ( nm ) | Relative STD (%) | LOD ( µg/acm ) |
|---|---|---|---|
| As | 278.02 | 9 | 600 |
| Be | 234.80 | 3 | 1 |
| Co | 345.35 | 8 | 24 |
| Cr | 425.44 | 5 | 7.8 |
| Cr | 359.30 | 5 | 12 |
| Zn | 330.30 | 15 | 570 |
| Cd | 228.80 | 5 | 45 |
| Hg | 253.65 | 13 | 680 |
| Sb | 259.81 | 9 | 120 |
| Sn | 283.99 | 10 | 190 |
| Mn | 257.61 | 4 | 4 |
| Mn | 403.08, 403.31, 403.45 | 8 | 7.5 |
| Ni | 341.48 | 9 | 30 |
| Pb | 405.78 | 6 | 90 |
| Fe | 404.58 | 6 | 140 |

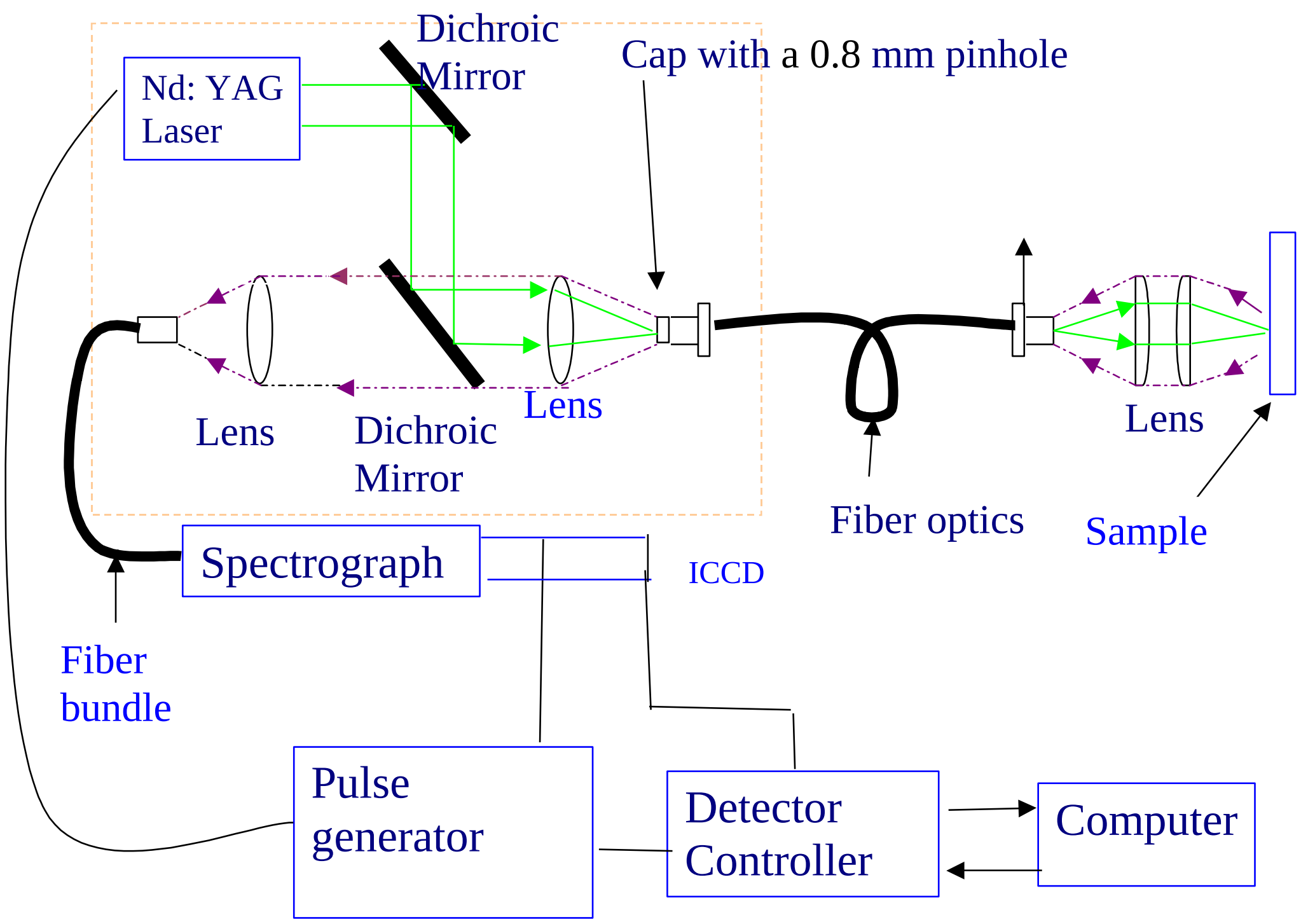

**Fig.-1 Schematic diagram of the FO LIBS system**

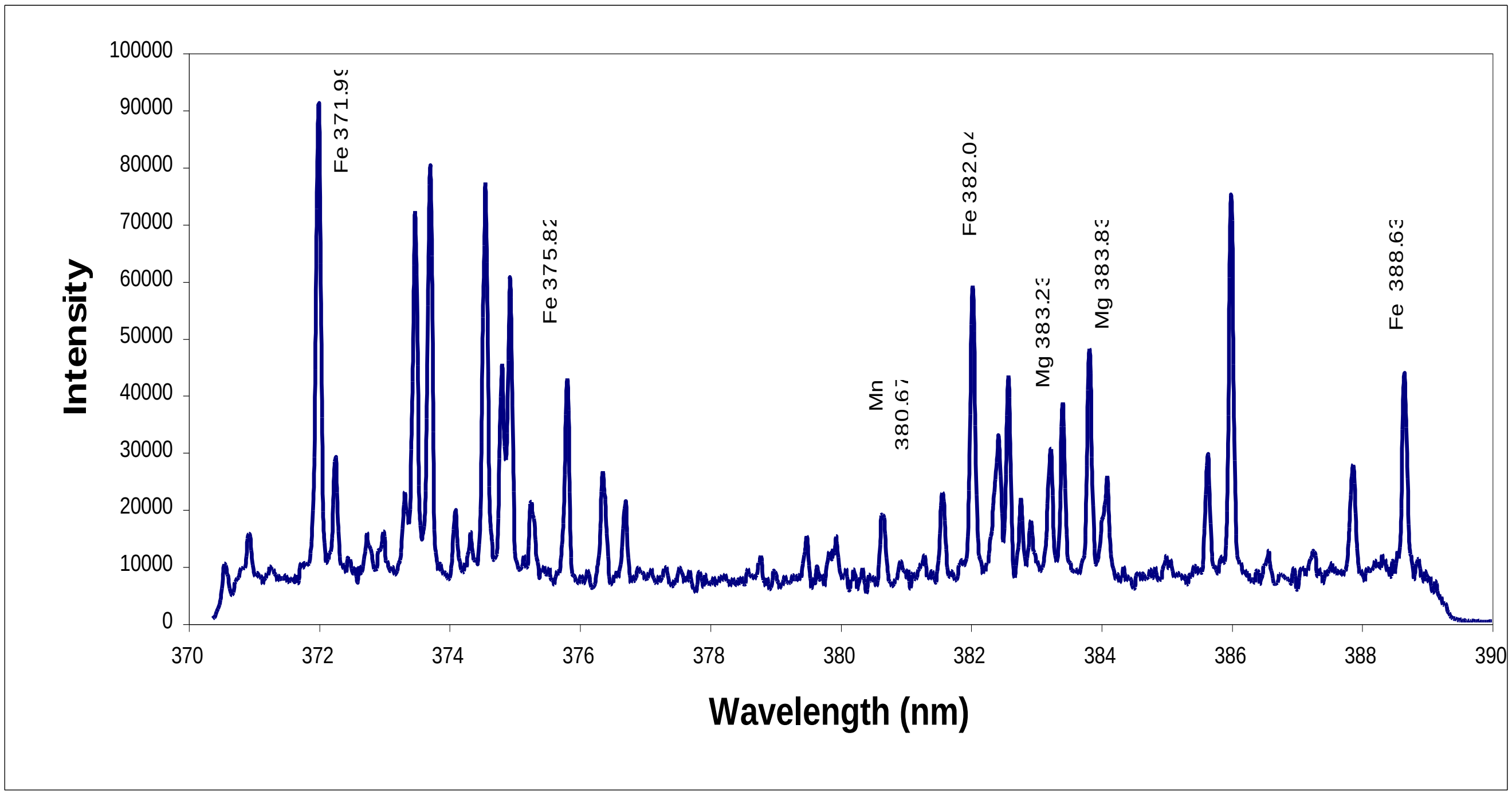

**Fig,-2 A typical LIBS spectrum of Al Alloy**

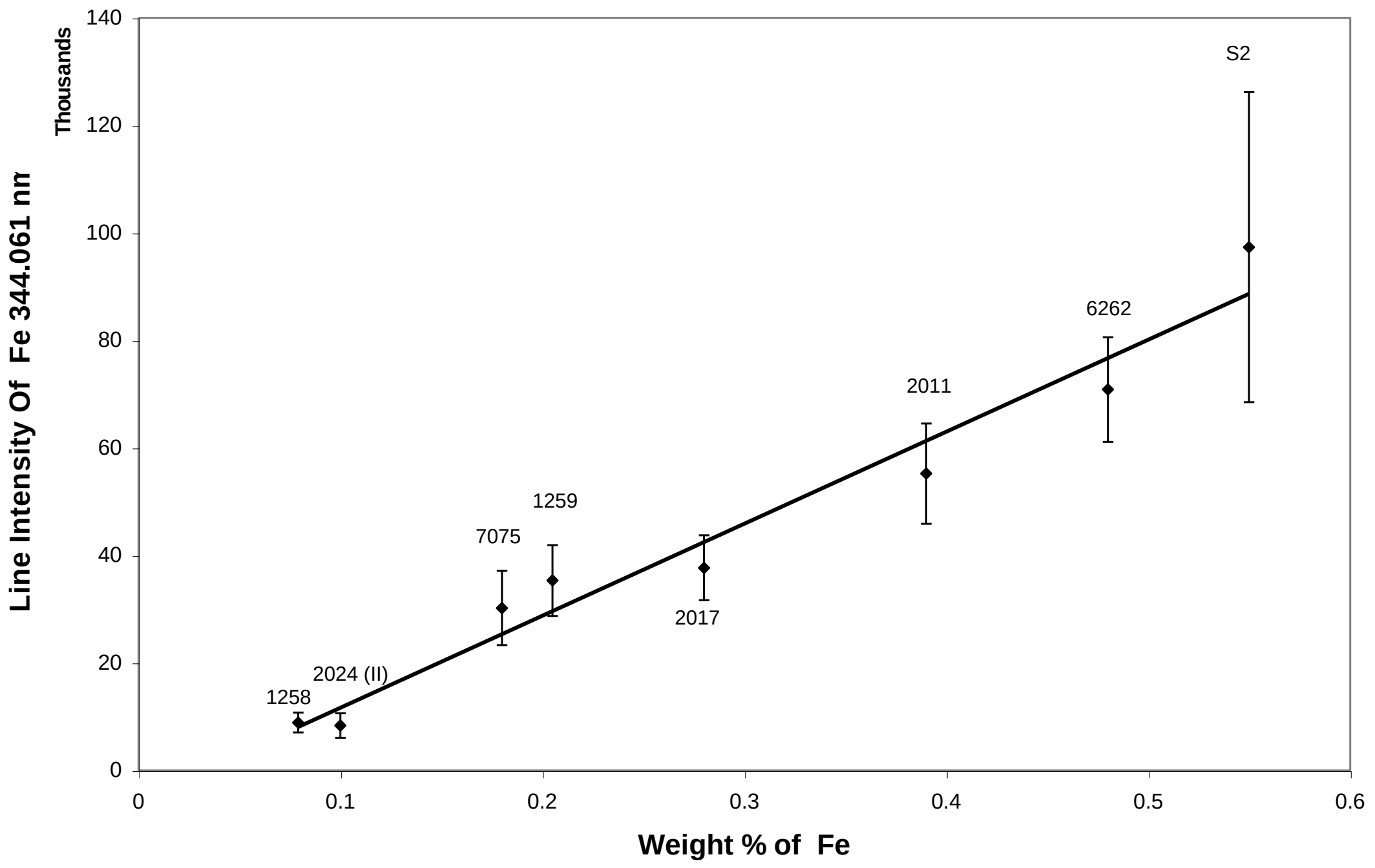

**Fig.-3 Calibration curve of Fe based on their absolute line intensity in LIBS spectra of solid Al alloy**

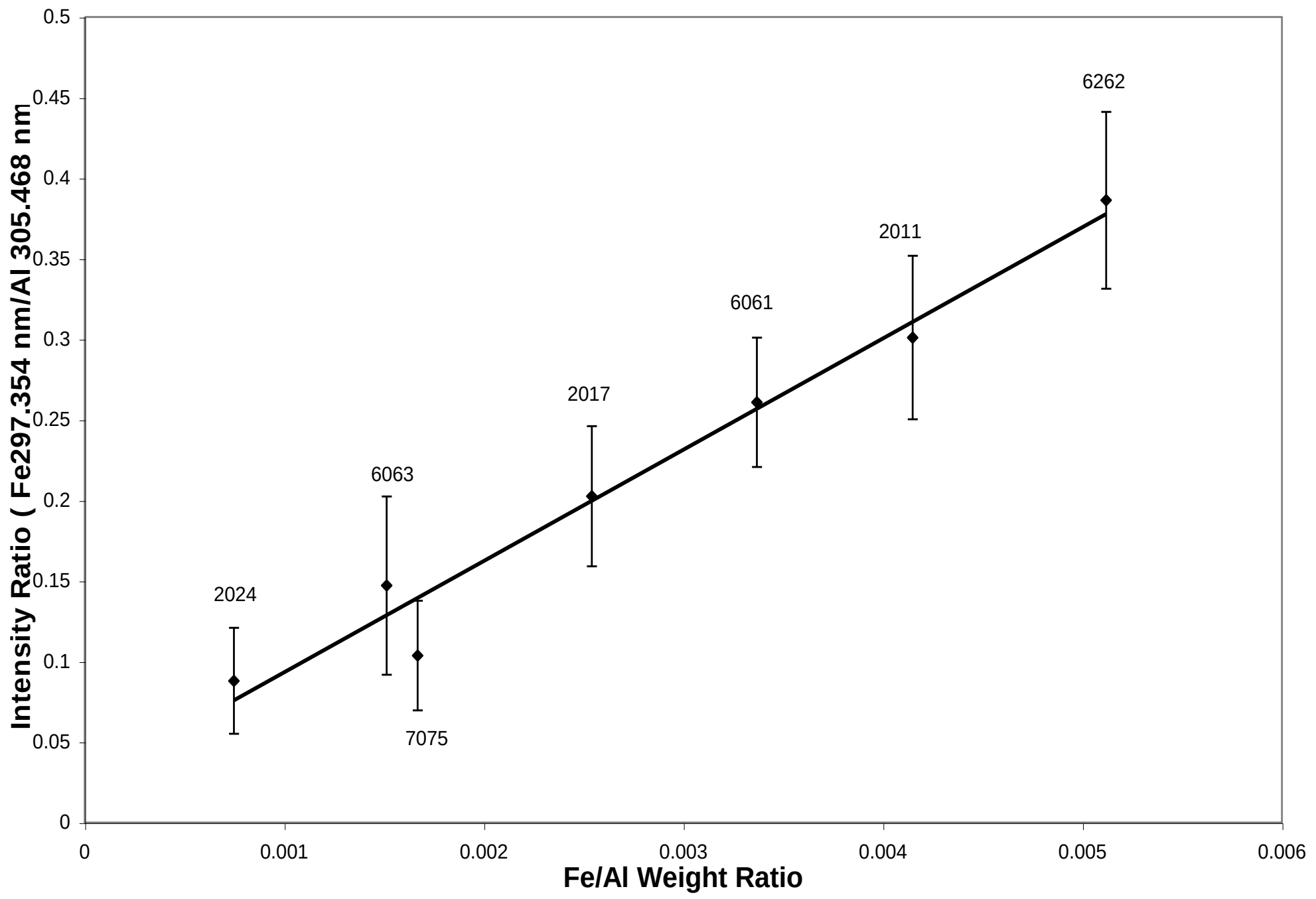

**Fig.- 4 Calibration curve of Fe using the ratio of an analyte line Fe with reference line Al in the LIBS spectra of solid Al alloy.**

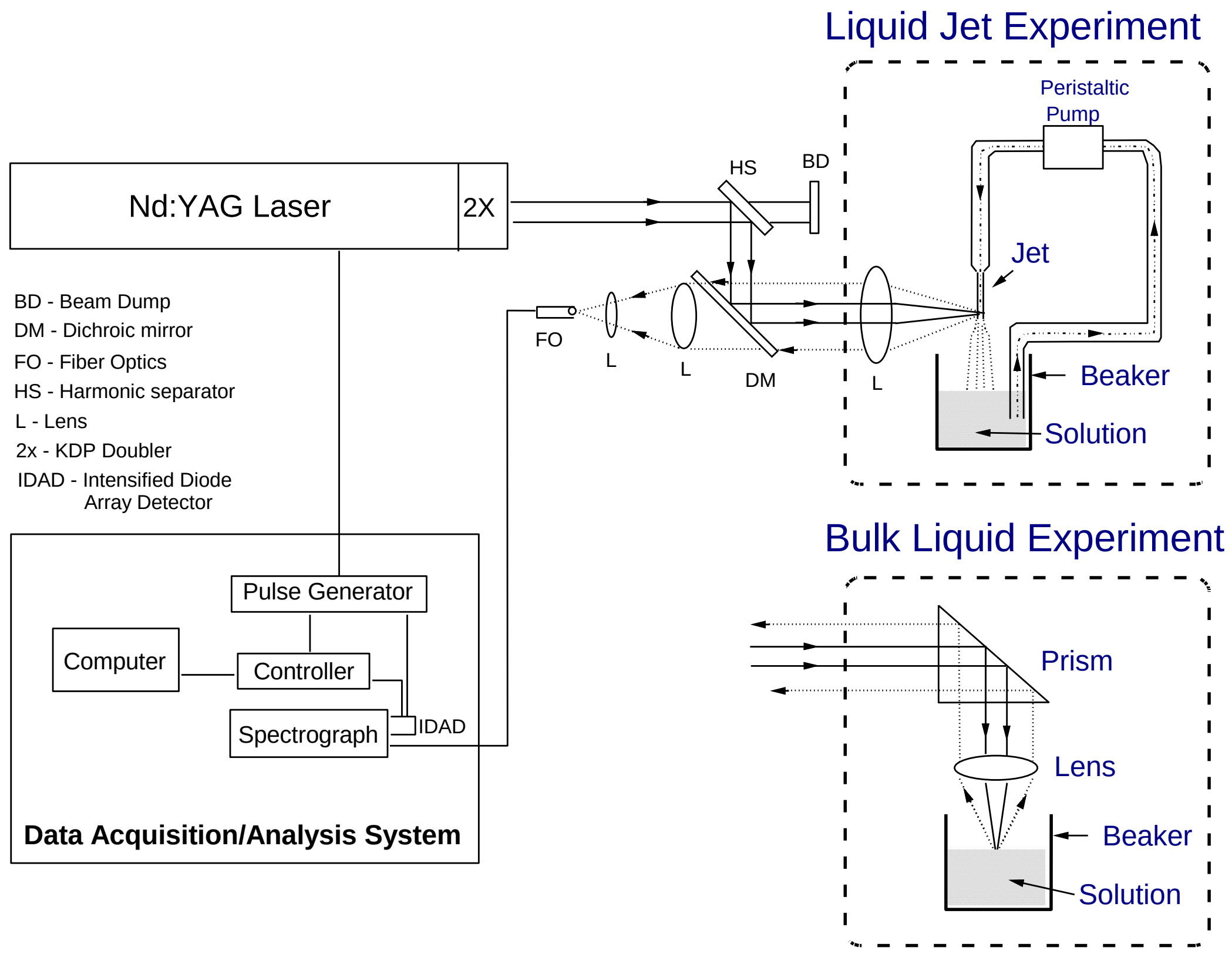

**Fig. –5 Experimental setup for recording laser-induced breakdown emission from Liquid.**

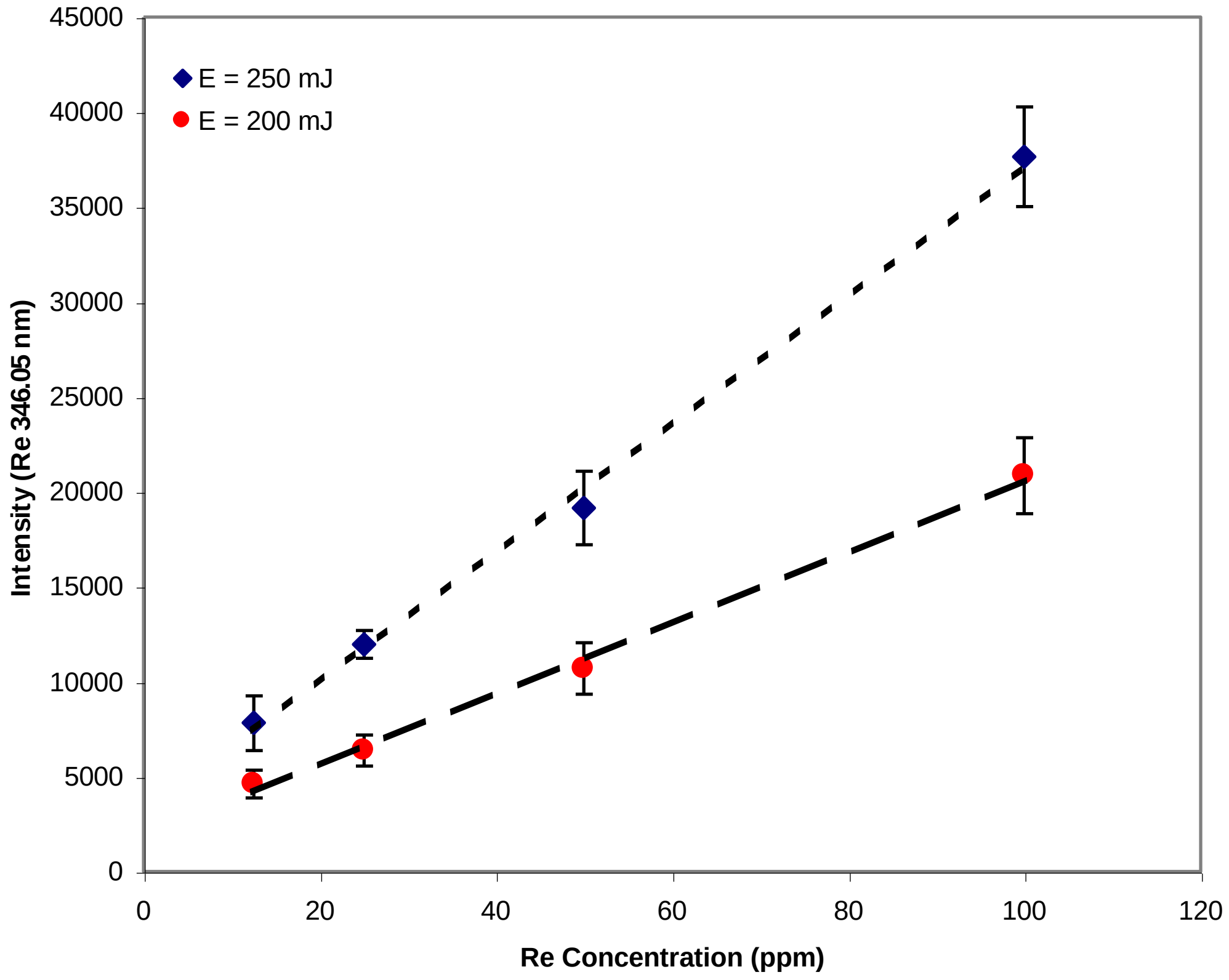

**Fig.-6 Calibration curve of Rhenium (Re) for different laser energy.**

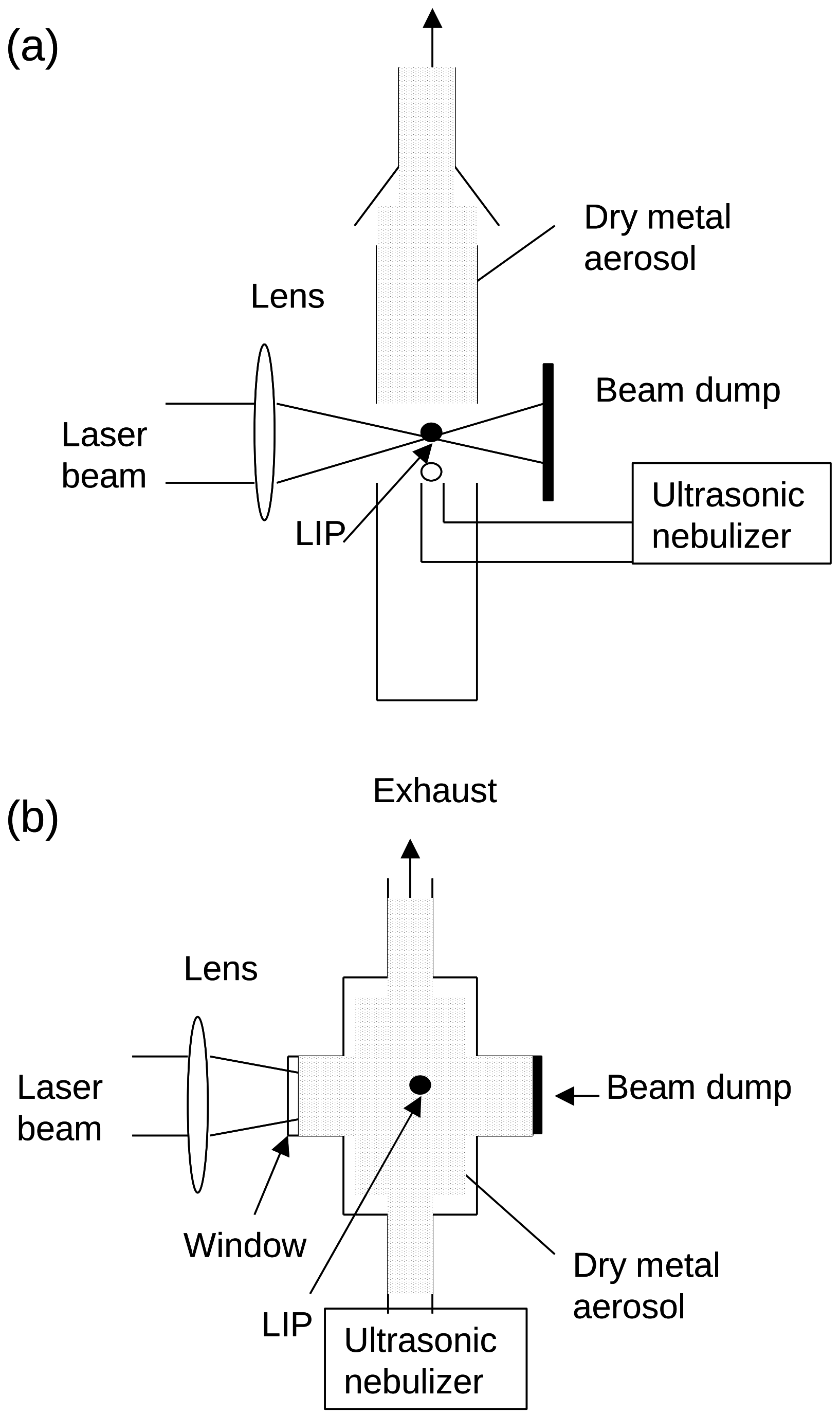

**Fig.-7 DIAL/LIBS calibration set up for gas sample (a) Open and (b) close system.**

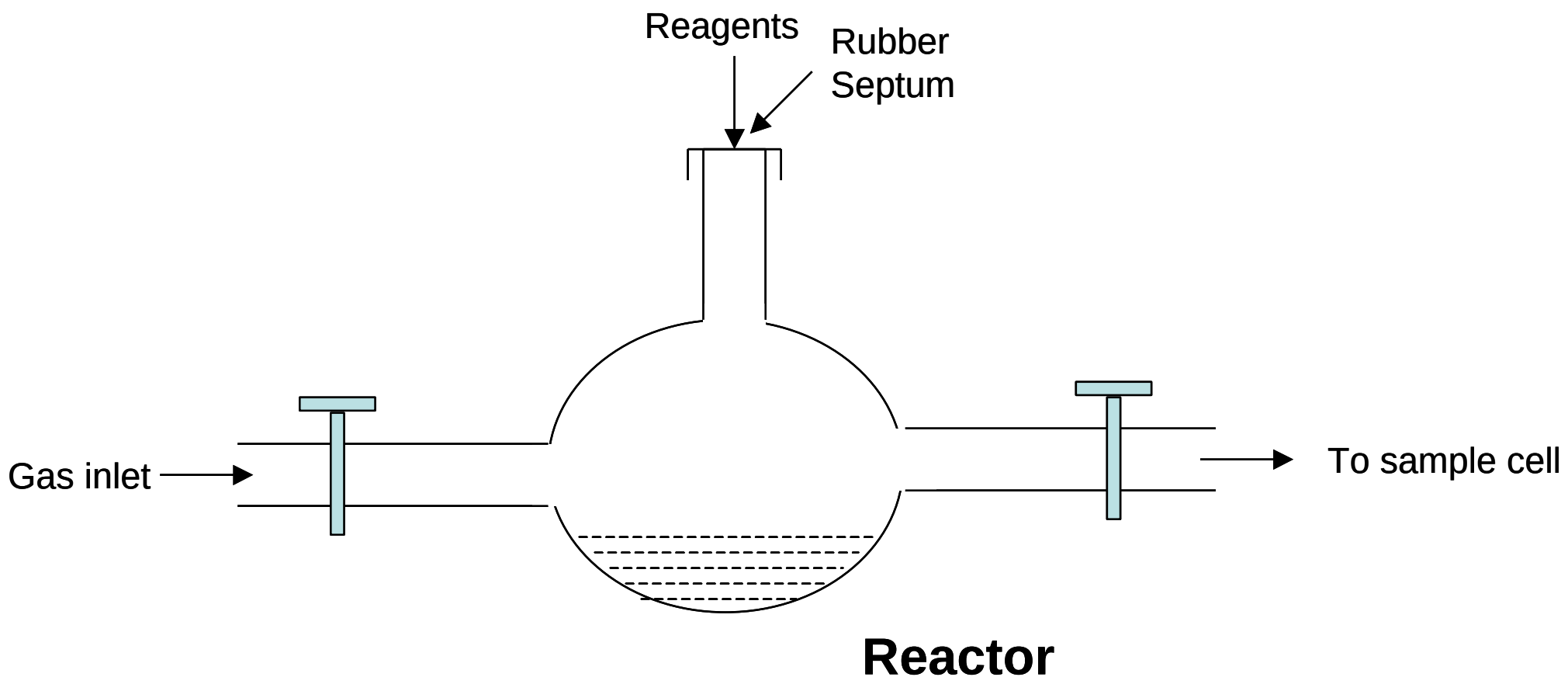

$$NaBH_4 + 3H_2O + HCl \longrightarrow H_3BO_3 + NaCl + 8H \xrightarrow{E^{m+}} EH_n + H_2(\text{excess})$$

**Fig.- 8 Schematic diagram of hydrite generator.**

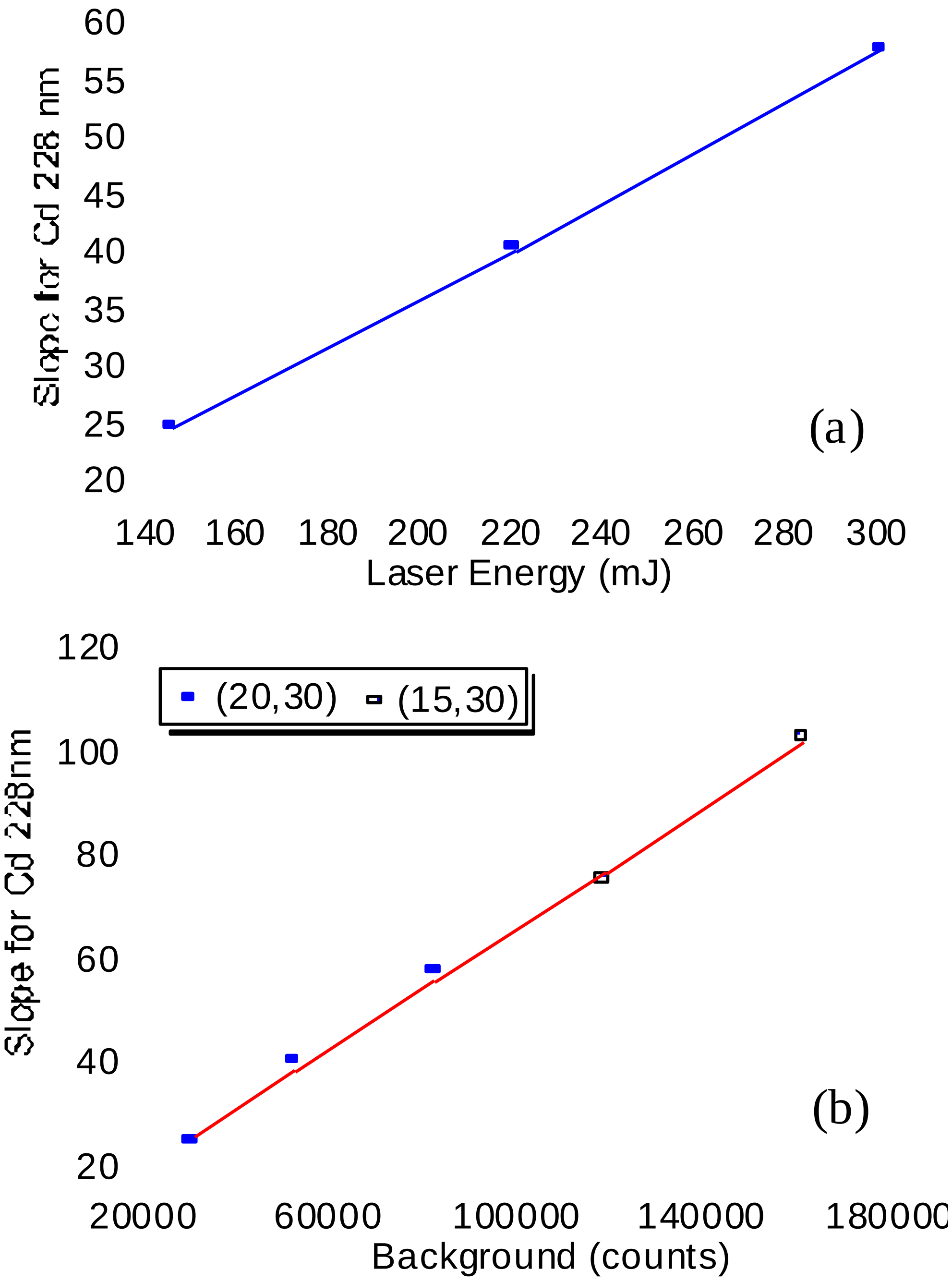

**Fig.-9 (a) Cd Calibration slope versus laser energy.**
**(b) Cd calibration slope versus LIBS background.**

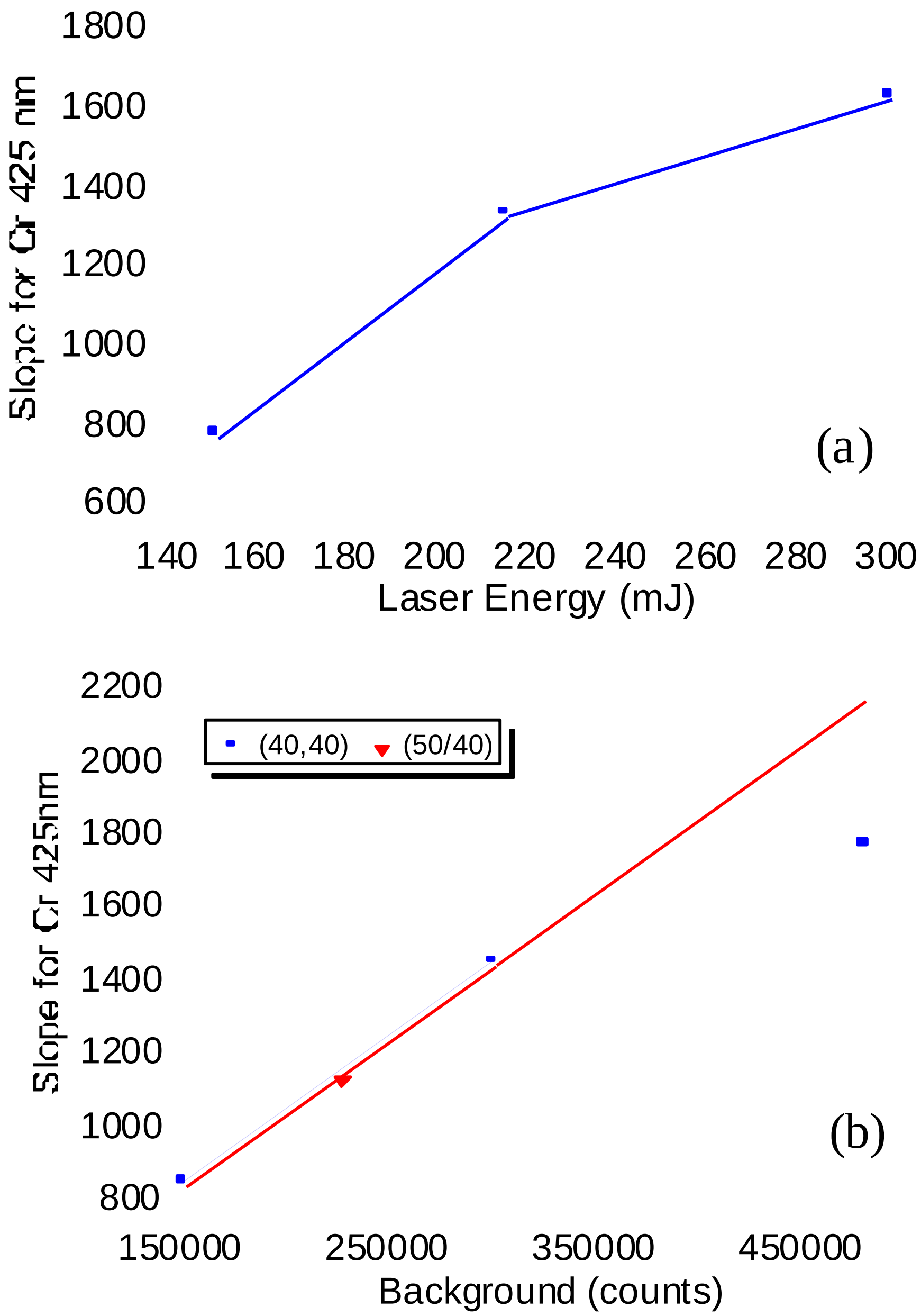

**Fig.-10 (a) Cr calibration versus laser energy.**
**(b) Cr Calibration slope versus LIBS background.**